\documentclass[twocolumn,showpacs,superscriptaddress,preprintnumbers,amsmath,amssymb]{revtex4}
\usepackage{graphicx} 
\usepackage{epsfig}
\usepackage{dcolumn} 
\usepackage{bm}
\usepackage{color}
\usepackage{times}
\usepackage{hyperref}
\usepackage{eurosym}
\begin{document}
\color{black}
\title{Error accounting in electron counting experiments}
\author{Michael Wulf}
\email{michael.wulf@ptb.de} 
\author{Alexander B. Zorin}
\affiliation{Physikalisch Technische Bundesanstalt, Bundesallee 100,
38116 Braunschweig, Germany}

\date{\today}

\begin{abstract}
%
Electron counting experiments attempt to provide a current of a
known number of electrons per unit time. We propose architectures
utilizing a few readily available electron-pumps or turnstiles with
the typical error rates of 1 part per $10^4$ with common sensitive
electrometers to achieve the desirable accuracy of 1 part in $10^8$.
This is achieved not by counting all transferred electrons but by
counting only the errors of individual devices; these are less
frequent and therefore readily recognized and accounted for. We
thereby ease the route towards quantum based standards for current
and capacitance.
\end{abstract}
 \maketitle
\begin{figure}
    \centering
    \includegraphics[width=3.5in, height=3.008in]{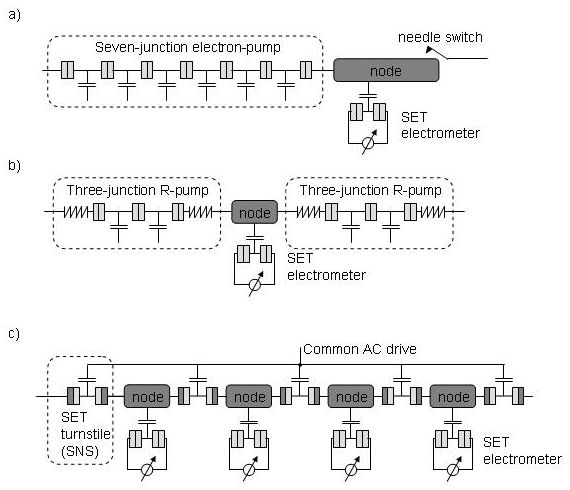}
    \caption{(a)
    Circuit schematic of an electron pump with seven tunnel junctions
    \cite{keller99}. The drives applied to the six intermediate islands
    are triangular pulses in carefully chosen phase-relationships. The electrometer on the final
    node is needed to verify device operation. A cryogenic needle switch
    makes contact to the chip to reversibly connect the pump to a cryo-capacitor.
    (b) Schematic of experiment with two sequential R-pumps and one intermediate node;
    resistors are drawn in red. The cryogenic switch can be removed from this device.
    (c) Schematic of proposed circuit of five single-gate pump devices, here based on the SNS-turnstiles of
    \cite{pekolaCurr}.
    These are connected in series with the intermediate nodes
    monitored
    by four electrometers. Capacitively coupled DC leads to the intermediate
    islands (not shown) allow adjustment of the required voltage-offsets of the SNS pumps, while
    all pump-gates are subjected to the same sinusoidal AC bias. }
    \label{fig:schem:pumpserrorcorrection}
\end{figure}

In 1990 the national metrological institutes adopted the Josephson
Volt \cite{JVolt,Kohlmann03} and von Klitzing Ohm \cite{Klitzing05}
as quantum based metrological units in addition to the conventional
SI units. Even earlier it had been realized \cite{met-tri,gallop05}
that a third electric quantum based standard would allow interesting
experiments to cross-check the theories describing Josephson and
quantum Hall effect. Particular interest has been spent on
developing a current standard based on counting all electrons
flowing in a given time interval through a constraint.
While a quantum current standard is highly desirable, an
experimentally more feasible alternative is an electron counting
based capacitance standard (ECCS): Using $N$ electrons to charge a
capacitor to a voltage $U$ allows to establish the devices
capacitance $C=Ne/U$. $U$ can be traced back to the Josephson
quantum voltage standard, while $e$ is the electron charge, thereby
$C$ is traced back to quantum units. As capacitance can also be
calculated for certain geometries \cite{CalcCap}, this were to
enable a comparison of two different ways to standardize
capacitance, but even for the ECCS the engineering challenges have
not yet been met in a reliable fashion, though a prototype has been
demonstrated \cite{keller99}. To surpass past achievement the
uncertainty in the number of counted electrons has to be reduced to
$10^{-8}$, complicating the task at hand. While capacitance
metrology requires only small currents, current metrology only
becomes possible if currents in excess 1 nA can be supplied
\cite{Keller2000}. This can be achieved by using pump frequencies
much higher than those that have shown satisfactory accuracy or by
parallelization of devices. A higher operation frequency increases
the error-rates significantly \cite{100MHzRpump}, while
parallelization efforts are hindered by conflicting requirements on
the devices, including yield and accuracy. We propose here a concept
to alleviate both concerns.
\\
The most advanced ECCS experiments use multi-stage
 devices referred to as electron-pumps, shown in fig. \ref{fig:schem:pumpserrorcorrection}(a). These utilize
successive quantum tunneling between metallic islands connected via
tunnel-junctions. Jensen and Martinis \cite{jensen92,jensen94}
recognized that electron pumps of several tunnel-junctions should be
able to provide a charge with metrologically relevant precision.
While relative errors of $10$ ppb have been reported in a 7-junction
electron pump \cite{keller99}, this scheme still requires continued
research to be robust enough as a metrological standard; device
yield is low, operation is complicated by the number of junctions
employed and the remaining error-rate is not theoretically
understood. The primary error in electron pumps is believed to be
caused by electron co-tunneling processes of order $N-1$, where $N$
is the number of junctions in the device \cite{jensen92}. These can
be suppressed by connecting resistors in series with the pump; the
resulting device is called the R-pump \cite{sergey01}. Other methods
to achieve this goal include the use of smaller or larger number of
tunnel junctions; both paths reduce yield while the latter also
complicates operation.
\\
Recently approaches based on simpler devices controlled by only a
single gate have received renewed attention. Quantized currents have
been observed in single-gate devices made in superconducting
\cite{delftQcurr} and in semiconducting technology
\cite{Kaestner07,Kaestner08a}, as well as in superconductor-metal
hybrid structures \cite{pekolaCurr}. Due to their operational
simplicity these approaches may be best suited for parallelization.
While these technologies may be operated also at frequencies higher
than those employed in the metallic electron-pump experiments,
thereby increasing available current, error-rates have not been
established for them and may prove to be higher than $10^{-8}$.
Recently  three-stage R-pumps have shown reasonable operation at
$100$ MHz as well \cite{100MHzRpump}.

{\it Architectures for discrete step error accounting} - Instead of
optimization of single devices, we propose to use two R-pumps (or
electron-pumps) with rather high relative error rates of
$2\Gamma^R=10^{-5}$ and sufficiently long hold times. A schematic
for the experiment is shown in fig.
\ref{fig:schem:pumpserrorcorrection}(b). The two R-pumps are
connected in series with an intermediate node, whose charge is
monitored by a sensitive electrometer, such as the RF-SET
\cite{RF-SET}. In this architecture the left pump starts by pumping
$\tilde{n}$ electrons onto the intermediate node with pump-frequency
$\Omega_P$. In addition to errors of this first pump, a charge may
tunnel through the entire second pump. The rate $2\lambda$ at which
this occurs is the inverse of the mean hold-time; the latter having
been reported to be about $10$ s for five-junction electron-pumps
\cite{jensen94} and $16$ s for the three-junction R-pump
\cite{sergey05}.
For simplicity we assume that all errors occur with equal chance for
either sign; then the probability $p(n)$ for $n$ charges to be on
the node after this first pump sequence is to leading order,
$p(\tilde{n}\pm1)\simeq\tilde{n}\Gamma^R(1-2\lambda\frac{\tilde{n}}
{\Omega_P})+\lambda\frac{\tilde{n}}
{\Omega_P}(1-2\tilde{n}\Gamma^R)\ll1$, which is dominated by the
first term for relevant parameters. After the first pump phase the
electrometer is used to measure the number of electrons on the node.
RF-SETs \cite{RF-SET} are among the suitable electrometers for this
purpose. Several laboratories have built RF-SETs with charge
resolution of about $\sigma_q=10^{-5} e/\sqrt{\textrm{Hz}}$
\cite{RF-SET,Bylander04} and even a ten-fold improvement over these
numbers has been demonstrated \cite{ref:RFset4K}. Typical bandwidths
exceed $10$ MHz, which is sufficient for the needs of this work. An
alternative readout is provided by inductively shunted SETs
\cite{Zorin2000,ref:AZ:LSET,ref:LSET}, which may achieve similar
charge resolution. 
A coupling capacitance of $C_C=100$ aF between the electrometer's
gate and the intermediate node is consistent with maintaining this
charge-resolution. If the nodes self-capacitance is
$C_{\textrm{node}}=2.5$ fF, the effective charge-resolution is only
reduced by a factor of $C_{\textrm{node}}/C_C=25$. With the
parameters stated, the probability of a measurement error using $10
\mu$s of integration time is less than
$10^{-9}$. 
$C_{\textrm{node}}$ is nonetheless large enough for stable operation
of the pumps as each electron will produce a voltage of only about
$60$ nV on this node. $\tilde{n}$ has to be kept below $\sim 1000$,
as the three-junction R-pump has a stable operation window of $100$
$\mu$V, with respect to the voltage across it \cite{sergey05}.
The measurement is followed by a pump-sequence of the second pump,
emptying the node into the second reservoir. Subsequently a second
measurement establishes the node´s charge state at the end of the
cycle, which thereafter can be repeated. If during either
measurement the charge on the island is found to deviate from the
value expected due to the design of pump-cycles, this deviation is
attributed to errors of the pump that was active since the previous
measurement. This assumption will be correct unless either, the
second pump experienced a hold error, or if one of the two
measurements was wrong. The probabilities for each sign being
$p^{\textrm{error}}_{\textrm{hold}}<(\tilde{n}/\Omega_P+2\tau_m)\lambda$
and
$p^{\textrm{error}}_{\textrm{measure}}=(1-erf(\frac{eC_C\sqrt{\tau_m}}{C_{\textrm{node}}\sigma_q}-.5))/2$.
Errors produced by the active pump are properly recognized and can
thus be accounted for, therefore the unaccounted errors no longer
scale with $\Gamma^R$.
 Depending on the
architecture, it may suffice to identify the errors made.
Alternatively, feedback mechanisms could be provided after
error-detection, for example by altering the number of pump-cycles,
to counter a detected error. Such feedback mechanisms will be more
device specific than the scope of the present work and are therefore
left out here. Three-junction R-pumps have been operated at
frequencies as high as $\Omega_P=100$ MHz \cite{100MHzRpump}. With
the parameters as stated and $\tilde{n}=1000$ the corrected relative
error-rate becomes
$\Gamma^R_{corr}=2(p^{\textrm{error}}_{\textrm{hold}}+p^{\textrm{error}}_{\textrm{measure}})/\tilde{n}=9\cdot10^{-9}$;
accordingly this proposed circuit reaches the metrologically
necessary accuracy.
\\
A further improvement may be achieved by use of a third pump and a
second intermediate island. The above method allows recognition of
the number of errors per cycle, which are then attributed to the
active device. With the additional island the signatures of the
errors for each of the three pumps differ: If an error occurs on the
left or right pump, only the charge on the corresponding island
deviates from expectation, while the charges on both islands do in
opposite directions whenever an error occurs on the middle pump. As
both nodes are monitored these errors can therefore be distinguished
and accounted for, though measurement errors remain. Furthermore,
errors involving two R-pumps may be indistinguishable from errors of
the remaining one. The probability of any of these errors occurring
during a complete cycle comprised of three pump- and
measurement-phases is approximated to the first order by $p_3 =
6(\tilde{n}/\Omega_P+2\tau_m)
\lambda(2\tau_m\lambda+\tilde{n}\Gamma^R) +
6p^{\textrm{error}}_{\textrm{measure}}$, and the corrected relative
error rate becomes $2\Gamma^R_{corr,3} = p_3/\tilde{n}$. Even
devices with performance values of $2\Gamma_R = 10^{-4}, \lambda =$
$1$ Hz, $\Omega_P = 100$ MHz suffice to provide the metrological
requirement, achieving $2\Gamma^R_{corr,3} = 6\cdot10^{-9}$ for
$\tilde{n}=53$ with an effective pump-rate reduced to $16.6$ MHz.
The operation even of three low-quality R-pumps in series is
expected to be experimentally less challenging than the operation of
a single near perfect device; the same is likely regarding
fabrication yield. In addition, the proposed architecture allows the
error-rates to be continuously monitored during operation, and
thereby in-situ minimization of error rates through
dynamic adjustment of external biases. %
This is not possible in the conventional ECCS experiments
\cite{Keller97}, as single electrons can no longer be detected on
the pumps final node once is connected to the external capacitor. In
the conventional ECCS experiment the electron-pump operation is only
optimized while the final node is isolated from the external
capacitor; then the electrometer allows detection of single
electrons on this node, allowing optimization by shuttling a few
electrons back and forth through the pump. Afterwards the pump is
connected to the capacitor by means of a needle switch contacting
onto this node. Misaligned needle-switches may damage devices;
furthermore the mechanical force inherent in the switching process
may shift the charge offsets on the islands due to the
piezo-electric effect \cite{KellerPrivComm}. In contrast, the
proposed architecture allows the switches removal from the chip
containing the pump-devices, and thereby further simplification of
the experiment.

{\it Architectures for continuous error accounting} -  The circuit
with two nodes and three pumping devices, allows further
improvement. Even a modest relative error-rate of $10$ ppm and an
operation frequency of $100$ MHz yields on average only $1000$
errors per pump per second. Such rare errors are readily resolved,
even while all three pumps operate continuously. Without pump-errors
the charge on the nodes were to remain constant, or only be
modulated at the pump frequency, while perfect and instant
measurements of the node charges allow for all pump-errors to be
detected and properly attributed according to the rules stated
above. But due to the finite charge-resolution of available
electrometers, there is a time $\tau_M$, during which errors
occurring, nearly simultaneously, in two of the pumps may be
indistinguishable from an error of the third device. The chance for
each of these six relevant second order scenarios to occur within
one measurement time is just $P_E=(\Omega_P\Gamma^R\tau_M)^2$. With
$\Omega_P=1$ MHz, $\Gamma^R=10^{-5}$ and $\tau_M=15$ $\mu$s we
recover $P_E=2.25\cdot10^{-8}$. If the information from the
measurements is then used to identify the single-pump-errors and
correct or account for them, the reduced, relative error rate
becomes of order
$2\gamma^R_3=6P_E/\tau_M\Omega_P=6(\Omega_P\Gamma^R\tau_M)\Gamma^R$
which computes to $2\gamma^R_3=9\cdot10^{-9}$, beating current
metrological requirements. When not three but $2N+1$ pumps are used,
the dominating error, that is not properly recognized, is of order
$N+1$ so that the leading term in the relative error rate scales as
\begin{equation}\label{eq:naiveerror}
2\gamma^R_{2N+1}=2\frac{(2N+1)!}{N!(N+1)!}(\Omega_P\Gamma^R\tau_M)^N\Gamma^R.
\end{equation}
Note that the error rate decreases exponentially with the number of
devices included. As both pump operation and measurements are
continuous, the following more complete approach is needed.
\\
For a circuit of $2N+1$ pumps with independent errors with rates
$2\alpha=2\Gamma^R\Omega_P$ with respect to time the joint
probabilities of $n_i$ errors having occurred at pump $i$ obey, with
$\vec{n}=\{n_1,n_2...,n_{2N+1}\}$, the n-dimensional
diffusion-equation
\begin{equation}\label{eq:probs:diffeeq}
\dot{P}_{\vec{n}}(t)=-2N\alpha
P_{\vec{n}}(t)+\sum_{|\vec{n}-\vec{m}|=1}\alpha
P_{\vec{m}}(t).
\end{equation}
The $(2N+1)$-dimensional set of probabilities, $P_{\vec{n}}$, can be
rearranged as a vector, $\vec{P}$. This allows rewriting of eq.
(\ref{eq:probs:diffeeq}) as $\dot{\vec{P}}(t)=T\vec{P}(t),$ defining
operator $T$. This diffusion corresponds to the errors in
conventional electron pumping experiments, while in our circuit this
diffusion is observed by the electrometers. The perceived
probabilities of the number of errors that have occurred will depend
on the diffusion of eq. (\ref{eq:probs:diffeeq}) as well as the
result of the measurement outcomes of the $N-1$ intermediate
electrometers, $\{x_1(t),..,x_{N-1}(t)\}$. In the absence of
diffusion and due to the measurements alone $\Pi_{\vec{n}}$, the
perceived probabilities of $\vec{n}$ errors having occurred divided
by the probability of $\vec{n_0}=\{0,0 .. 0\}$ errors having
occurred, develop according to
\begin{equation}
\dot{\Pi}_{\overrightarrow{n}}=[\sum_{i=1}^{N-1}(2(n_i-n_{i+1})x_i-(n_i-n_{i+1})^2)/\beta_i]\Pi_{\overrightarrow{n}}.
\end{equation}
Here $\beta_i$ denotes the rms-mean of the noise of the electrometer
$i$, normalized by the single-electron signal, which is assumed to
be gaussian and white. This equation can be rewritten to yield
$\dot{\overrightarrow{\Pi}}(t)=U(t)\overrightarrow{\Pi}(t).$ Observe
that $U$ is stochastic and includes information about the
instantaneous electrometer signal, which in turn is a function of
the instantaneous electrometer noise as well as the diffusion
history.
\\
In the scenario of interest both processes coexist; from the
measured traces of the electrometer signals we extract the evolution
of the perceived relative probabilities, $\overrightarrow{\pi}$, as
\begin{equation}\label{eq:eqom:completeprobs:operators}
\dot{\vec{\pi}}=(T+U)\vec{\pi}.
\end{equation}
\begin{figure}
    \centering
    \includegraphics[width=3.5in, height=1.574in]{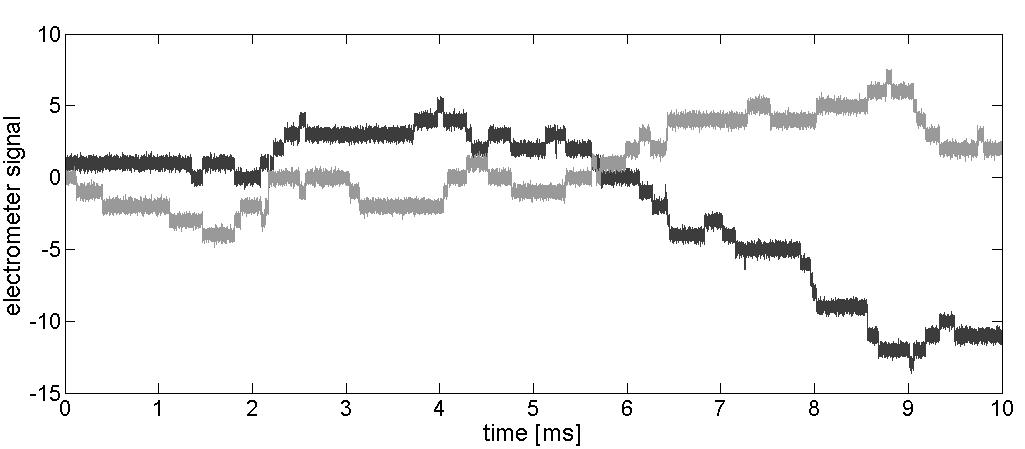}
    \caption{Signal traces of first (black) and second (gray) electrometer.
    Black and red lines indicate ideal measurement results without noise. Pump-error
    rates are $1000/$s, per pump and sign, electrometer noise $10^{-5}e/\sqrt{\textrm{Hz}}$, coupling capacitance
    ratio $C_C/C_{\textrm{node}}=0.04$, signal averaged over $200$ ns.}
    \label{fig:b2a:sim:signaltraces}
\end{figure}
Figure (\ref{fig:b2a:sim:signaltraces}) shows a typical set of
signal traces that can be expected from the combination of the
diffusion dynamics and likely electrometer noise. These traces are
used to numerically integrate eq.
(\ref{eq:eqom:completeprobs:operators}) yielding a perceived
diffusion. Figure \ref{fig:b2a:sim:histogram} contains a histogram
of the result of $12500$ generated diffusions contrasted with the
effective error resulting when for each trace the perceived
diffusion is subtracted from the underlying diffusion, yielding the
error despite correction.
\begin{figure}
    \centering
    \includegraphics[width=3.5in, height=1.574in]{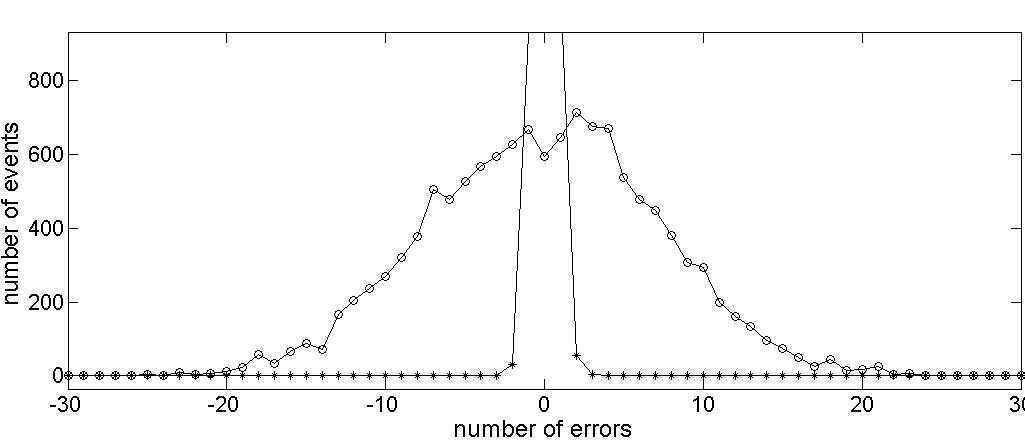}
    \caption{Histogram of the errors (circles) and errors after correction (stars) of $12500$ traces
    of $30$ ms with an error rate of $1000/$s, other parameters as in fig.
    (\ref{fig:b2a:sim:signaltraces}). The peak corresponding to proper recognition of all
    errors is
    suppressed for clarity.}
    \label{fig:b2a:sim:histogram}
\end{figure}
\begin{figure}
    \centering
    \includegraphics[width=3.5in, height=1.574in]{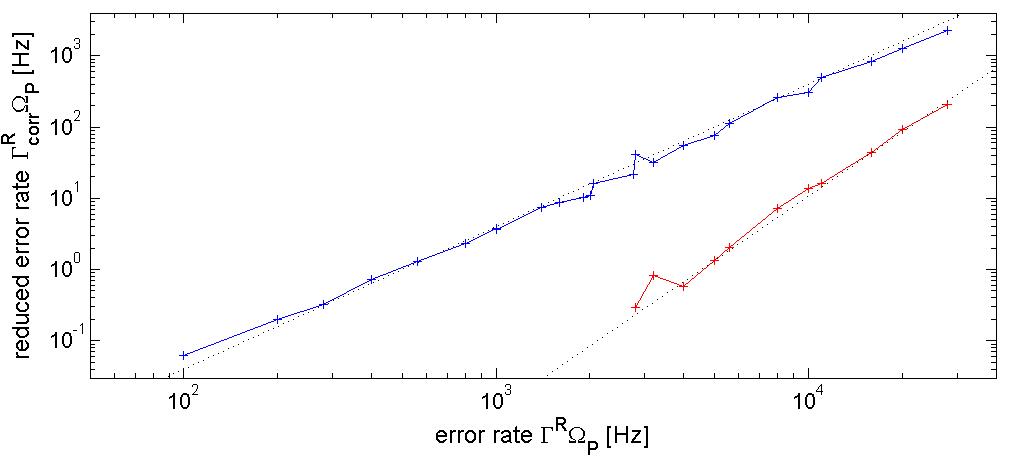}
    \caption{Corrected error rates with respect to time for three (upper curve) and five (lower curve) pump
    circuits as a function of the uncorrected error rates. Dashed
    lines are a fit to eq. (\ref{eq:naiveerror}). Other
    parameters as in fig. \ref{fig:b2a:sim:signaltraces}.}
    \label{fig:b2a:sim:ErrorRates}
\end{figure}
The latter corresponds as well to a diffusion, albeit with a reduced
diffusion (or error) rate. Numerical results for these reduced error
rates are shown for various parameters in fig.
\ref{fig:b2a:sim:ErrorRates} in excellent agreement with eq.
(\ref{eq:naiveerror}). Note that even for a rate of $200$ errors per
second a $10000$-fold improvement of the error rate is achieved with
only three pumps and two intermediate electrometers, so that even
devices exhibiting a relative error-rate of $100$ ppm can reach the
metrologically desirable error threshold of $10$ ppb due to the
proposed architecture. 
 The five pump version achieves a similar improvement
even at a rate of $8000$ errors per second. For devices with a
relative error-rate of $100$ ppm, this would allow pump rates of 2
MHz for three devices or $80$ MHz for five. If however devices with
a relative error-rate of $10$ ppm were available, only a
thousandfold improvement were required; this could be achieved with
a five-pump architecture and a pump frequency of approximately $1$
GHz. The latter example not only achieves the required accuracy for
a metrological current standard, but achieves this with sufficiently
large current for applications such as cross-checking the relations
of the metrological triangle of
electrical units. \\
In summary, we have proposed a family of architectures utilizing
either conventional electron pumps of modest quality or one of the
single-gate devices that are capable of producing quantized current
steps with the now common RF-SET or a similar device. The algorithm
employed in this work determines the charge transported by such
structure to a much greater precision than would be feasible by any
single of the devices used. The relaxation on the requirements for
the quality of the individual devices used, simplifies
experimentation, even at higher operation frequency, and increases
sample yield. It thereby opens the prospects towards parallelization
needed to increase the provided current. We note that the proposed
architecture allows to monitor the error-rates of individual pumps
on the level of single electrons, similar to those stated for
electron-pumps in their shuttle mode. This has not been possible for
the single gate devices \cite{delftQcurr,Kaestner07,pekolaCurr} in
currently employed circuits. Accordingly the architecture proposed
here allows further research on these devices and thereby a
determination regarding their suitability for metrological
applications, while easing the required thresholds.

This work is supported in part by the European Union under the Joint
Research Project REUNIAM as well as through the project EuroSQIP. M.
W. gratefully acknowledges discussions with  M. J. Feldman on the
early stages of this work.


\begin{references}

\bibitem{JVolt}
M. T. Levinson, R. Y. Chia, M. J. Feldman and B. A. Tucker,
\newblock {\em Appl. Phys. Lett.} {\bf31}, 776 (1977).

\bibitem{Kohlmann03}
J. Kohlmann, R. Behr and T. Funck,
{\em Meas. Sci. Technol.} {\bf14}, 1216 (2003).

\bibitem{Klitzing05}
K. von Klitzing,
\newblock {\em Phil. Trans. R. Soc. A} {\bf363}, 2203 (2005).

\bibitem{met-tri}
K. K. Likharev and A. B. Zorin
\newblock {\em Jour. Low Temp. Phys.} {\bf59}, 347 (1985).

\bibitem{gallop05}
J. Gallop,
\newblock {\em Phil. Trans. R. Soc. A} {\bf 363}, 2221 (2005).

\bibitem{CalcCap}
W. K. Clothier,
\newblock {\em Metrologia} {\bf1}, 36 (1965).

\bibitem{keller99}
M. W. Keller, A. L. Eichenberger, J. M. Martinis, N. M. Zimmermann,
\newblock {\em Science} {\bf285}, 1706 (1999).

\bibitem{Keller2000} M. W. Keller,
\newblock in {\it proceedings of Fermi school CXLVI: Recent
Advances in Metrology and Fundamental Constants} (2001).

\bibitem{100MHzRpump}
\newblock B. Steck, A. Gonzalez-Cano, N. Feltin, L. Devoille, F. Piquemal,
S. Lotkhov and, A. B. Zorin
\newblock {\em Metrologia} {\bf45}, 482 (2008).

\bibitem{jensen92}
\newblock H. D. Jensen and J. M. Martinis,
\newblock {\it Phys. Rev. B} {\bf 46}, 13407 (1992).

\bibitem{jensen94}
\newblock J. M. Martinis, M. Nahum and H. D. Jensen,
\newblock {\it Phys. Rev. Lett.} {\bf72}, 904 (1994).

\bibitem{sergey01}
S. V. Lotkhov, S. A. Bogoslovsky,A. B. Zorin and J. Niemeyer,
\newblock {\em Appl. Phys. Lett.} {\bf78}, 946 (2001).

\bibitem{delftQcurr}
\newblock L. J. Geerligs et al.
\newblock {\em Phys. Rev. Lett.} {\bf64}, 2691 (1990).

\bibitem{Kaestner08a}
\newblock B. Kaestner, V. Kashcheyevs, G. Hein, K. Pierz, U.
Siegner, and H. W. Schumacher,
\newblock {\em Appl. Phys. Lett.} {\bf92},
192106 (2008).

\bibitem{Kaestner07}
\newblock M. D. Blumethal, B. Kaestner, L. Li, S. Giblin, T. J. B. M.
Janssen, M. Pepper, D. Anderson, G. Jones, and D. A. Ritchie,
\newblock {\em Nature Physics} {\bf3}, 343 (2007).

\bibitem{pekolaCurr}
J. P. Pekola, J. J. Vartiainien, M. Möttönen, O.-P. Saira, M.
Meschke, and D. V. Averin,
\newblock {\em Nature Physics} {\bf4}, 120 (2007).

\bibitem{RF-SET}
R. J. Schoelkopf, P. Wahlgren, A. A. Kozhevnikov, P. Delsing, D. E.
Prober,
\newblock{\em Science} {\bf280}, 1238 (1998). 

\bibitem{sergey05}
H. Scherer, S. V. Lotkhov, G.-D. Willenberg, and A. B. Zorin,
\newblock {\em IEEE Trans. on Instr. and Meas.} {\bf54}, 666 (2005).

\bibitem{Bylander04}  J. Bylander, T. Duty and P. Delsing,
\newblock {\em Nature} {\bf434}, 361 (2005). 

\bibitem{ref:RFset4K}
H. Brenning, S. Kafanov, T. Duty, S. Kubatkin and P. Delsing,
\newblock {\em Jour. Appl. Phys.} {\bf100}, 114321 (2006).

\bibitem{Zorin2000}
A. B. Zorin,
\newblock {\em Phys. Rev. Lett.} {\bf86}, 3388 (2001).

\bibitem{ref:LSET}
M. A. Sillanp\"{a}\"{a}, L. Roschier, P. J. Hakonen,
\newblock {\em Phys. Rev. Lett.} {\bf 93}, 066805 (2004).

\bibitem{ref:AZ:LSET}
A.B. Zorin
\newblock {\em Physica C} {\bf 368}, 284 (2002).

\bibitem{Keller97}
M. W. Keller, J. M. Martinis, A. H. Steinbach, and N. M. Zimmerman,
\newblock {\em IEEE Tran. Instr. and Meas.} {\bf 46}, 307 (1997).

\bibitem{KellerPrivComm}
M. W. Keller, private communications.

\end{references}
\end{document}